\documentclass[showpacs,10pt,twocolumn]{revtex4-1}
\pdfoutput=1
\usepackage[hyperindex,breaklinks]{hyperref}
\usepackage{atlasphysics}
\usepackage{amsmath}
\usepackage{subfigure}
\usepackage{comment}
\usepackage{graphicx}
\usepackage{preprintcover}
\PreprintCoverPaperTitle{Search for Higgs Boson Pair Production in the $\gamma\gamma b\bar{b}$ Final State \\ using $pp$ Collision Data at $\sqrt{s}=8\tev$ from the ATLAS Detector}

\PreprintIdNumber{CERN-PH-EP-2014-113}
\PreprintCoverAbstract{Searches are performed for resonant and non-resonant Higgs boson pair production 
  in the {\ensuremath{\gamma\gamma b \overline b}} final state using 20\ensuremath{\text{~fb\tss{-1}}} of proton--proton collisions at a center-of-mass energy of 8~TeV
  recorded with the ATLAS detector at the CERN Large Hadron Collider. 
A 95\% confidence level upper limit on the cross section times branching ratio of non-resonant production 
  is set at 2.2~pb, while the expected limit is 1.0~pb. The difference derives from a modest excess of events, corresponding to 2.4 standard deviations from the background-only hypothesis.
The limit observed in the search for a narrow {\ensuremath{ X\to hh}} resonance
  ranges between 0.7 and 3.5~pb as a function of the resonance mass.
}

\PreprintJournalName{PRL}

\newcommand{\pb}{\ensuremath{\text{~pb}}}

\renewcommand{\ifb}{\ensuremath{\text{~fb\tss{-1}}}}
\newcommand{\tss}[1]{\textsuperscript{#1}}
\newcommand{\hhyybb}{\ensuremath{hh \rightarrow \gamma\gamma b \overline b} }
\newcommand{\htohh}{\ensuremath{H \rightarrow hh} }
\newcommand{\Hhh}{\htohh}

\newcommand{\yybb}{\ensuremath{\gamma\gamma b\overline{b}}}

\newcommand{\myybb}{\ensuremath{m_{\gamma\gamma b\overline{b}}}}
\newcommand{\myyjj}{\ensuremath{m_{\gamma\gamma jj}}}
\newcommand{\myy}{\ensuremath{m_{\gamma\gamma}}}
\newcommand{\mbb}{\ensuremath{m_{b\overline{b}}}}
\newcommand\hbb{\ensuremath{\Hboson \rightarrow b \overline b} }
\newcommand{\progname}[1]{\textsc{\texttt{#1}}}
\newcommand{\unsim}{\raisebox{.25ex}{\ensuremath{\mbox{\scriptsize $\sim\,$}}}}
\newcommand\hgg{\ensuremath{\Hboson \rightarrow \gamma\gamma} }
\newcommand\hyy{\hgg}

\newcommand\tth{\ensuremath{\ttbar\Hboson} }

\renewcommand{\Hboson}{\ensuremath{h}}

\begin{document}

\title{Search for Higgs Boson Pair Production in the $\gamma\gamma b\bar{b}$ Final State \\ using $pp$ Collision Data at $\sqrt{s}=8\tev$ from the ATLAS Detector}
\begin{abstract}
Searches are performed for resonant and non-resonant Higgs boson pair production 
  in the {\ensuremath{\gamma\gamma b \overline b}} final state using 20\ensuremath{\text{~fb\tss{-1}}} of proton--proton collisions at a center-of-mass energy of 8~TeV
  recorded with the ATLAS detector at the CERN Large Hadron Collider. 
A 95\% confidence level upper limit on the cross section times branching ratio of non-resonant production 
  is set at 2.2~pb, while the expected limit is 1.0~pb. The difference derives from a modest excess of events, corresponding to 2.4 standard deviations from the background-only hypothesis.
The limit observed in the search for a narrow {\ensuremath{X\to hh}} resonance
  ranges between 0.7 and 3.5~pb as a function of the resonance mass.
\end{abstract}
\maketitle

Within two years of discovering a new boson with a mass near $125\gev$ \cite{Higgs:120917_atlas_discovery,Higgs:120917_cms_discovery},
the ATLAS and CMS collaborations have completed a slate of measurements 
demonstrating that its spin and couplings conform to the predictions of the
Standard Model (SM) Higgs boson within current experimental and theoretical uncertainties~\cite{Higgs:131007_coupling_comb,Chatrchyan:2013mxa}.
Despite the lack of deviations from SM predictions, the Higgs boson, $h$, offers a rich potential for new physics searches.
This Letter reports on searches for non-SM physics 
  with events consistent with either resonant ($X\rightarrow hh$) or non-resonant pair production of Higgs bosons in the \hhyybb{} channel.

The predicted rate for Higgs boson pair production in the SM 
  is several orders of magnitude smaller than the rate 
  for the single $h$ process~\cite{Dawson:1998py,Grigo:2013rya,deFlorian:2013jea,Dihiggs:121221_djouadi_self-coupling}; 
  $hh$ production is thus not expected to be observable with current LHC data sets. 
However, a variety of extensions to the SM predict an enhancement of Higgs boson pair production.
In two Higgs doublet models (2HDMs) \cite{Books:050503_djouadi_anatomy_2,Dihiggs:140305_baglio_2hdm_benchmarks,Dihiggs:130510_craig_2hdm}
  the heavier of the neutral scalar Higgs bosons, $H$, may decay to a pair of its lighter scalar partners, $h$.
Depending on the parameters of the 2HDM, the $\Hhh$ production cross section
  may exceed a picobarn~\cite{Dihiggs:130510_craig_2hdm}. 
A deviation of the Higgs boson self-coupling, $\lambda_{hhh}$,
  from the SM predicted value could also increase the non-resonant production rate. 
Such deviations could be observed with future data sets~\cite{Dihiggs:121221_djouadi_self-coupling}.
Larger enhancements in the $pp \to hh$ rate could arise from the top-Higgs quartic $t\bar{t}hh$ coupling predicted in composite models~\cite{Grober:2010yv,Dihiggs:120524_contino_anomalous_hh},
  or from the addition of light colored scalars to the SM~\cite{Dihiggs:120718_kribs_light_scalars_hh}.
Resonant production of two Higgs bosons could appear from the production and decay of gravitons, radions or stoponium~\cite{Dihiggs:120731_gravitons,Dihiggs:001204_radions, Dihiggs:140403_martin_stop},
  as well as from a hidden sector mixing with the observed Higgs boson~\cite{Dihiggs:121030_dolan_englert_new_physics}.

The \yybb{} channel is an excellent final state for a search for Higgs boson pair production \cite{Dihiggs:031003_baur_self_coupling}
  thanks to the large $\hbb$ branching ratio, clean diphoton trigger, excellent diphoton invariant mass resolution, and low backgrounds.
This channel is particularly important in the search for resonances with mass, $m_X$, in the range $260 < m_X < 500\gev$ considered in this Letter, 
  where backgrounds and combinatorics make other channels such as $b\overline{b}b\overline{b}$ or $b\bar b\tau^+\tau^-$ challenging.

Processes that do not contain Higgs bosons are estimated from data; 
  all other processes are simulated using Monte Carlo techniques. 
The standard ATLAS detector simulation~\cite{Generators:100520_atlas_simulation_inf} based on \progname{Geant4}~\cite{Generators:020809_geant4} is used.
The simulation parameters are tuned to describe soft components of hadronic final states~\cite{Generators:120822_atlas_pythia8, Generators:111121_atlas_further_tunes}.
Simulated minimum bias collisions are overlaid on the hard scatter process, 
  and events are reweighted so that the average number of interactions per bunch crossing ($\unsim20$) matches the observed distribution.

Background events with a single Higgs boson produced in association with a $W$ or $Z$ boson or $\ttbar$ ($Wh$, $Zh$, $\tth$)
  are simulated with \progname{Pythia8}~\cite{Generators:071020_pythia8} using CTEQ6L1 
  parton distribution functions of the proton (PDFs)~\cite{Nadolsky:2008zw}.
Higgs boson production via gluon or vector-boson fusion (ggF, VBF)
is simulated using CT10 PDFs~\cite{Lai:2010vv} with \progname{Powheg-Box}~\cite{Generators:090115_powheg_ggF,Generators:091211_powheg_VBF} interfaced to \progname{Pythia8} for showering and hadronization.
Cross sections and associated uncertainties are taken from Ref.~\cite{Higgs:130704_YR_3_prop}.

Two benchmark signal models are defined: 
  SM Higgs boson pair production for the non-resonance search, 
  and a gluon-initiated, spin-0 resonant state in the narrow-width approximation for the resonance search. The SM $hh$ process is too small to observe with current data sets, but the SM kinematics are used to model generic non-resonant beyond-SM physics. 
Both models are generated using \progname{MadGraph5}~\cite{Generators:110602_madgraph5,Generators:131212_maltoni_madraph_pair_prod} and CTEQ6L1 PDFs.
A generator filter requires a pair of $b$-quarks and a pair of photons in each event.
\progname{Pythia8} is used to decay the two Higgs bosons,
  and to shower and hadronize the events.
The implementation of SM Higgs boson pair production includes the interference between diagrams with trilinear Higgs boson couplings and box diagrams. For the SM $hh$ process, which is a background to the resonance search, the next-to-leading-order inclusive production cross section of $9.2$~fb 
  is taken from Ref.~\cite{Dihiggs:121221_djouadi_self-coupling}.
Resonant samples are generated with a width of $10\mev$ (corresponding to a narrow width approximation) at masses $m_X = 260,\,300,\,350$, and 500~GeV.
Production cross sections for benchmark 2HDM models are calculated with \progname{SusHi}~\cite{Harlander:2012pb}, and branching ratios with \progname{2HDMC}~\cite{Eriksson:2009ws}.

The analysis described in this Letter uses the full $\sqrt{s}=8\tev$ data set of proton--proton collisions recorded by the ATLAS experiment in 2012, corresponding to an integrated luminosity of $20.3\pm 0.6$\ifb{}~\cite{Detectors:130218_atlas_lumi}. 
Data quality criteria are applied to reject events with diminished detector performance~\cite{Aad:2014una,dq}. 
A description of the ATLAS detector can be found elsewhere~\cite{Detectors:080108_atlas_detector}.

The photon and event selection for the present search largely 
follow those of published ATLAS $h \rightarrow \gamma\gamma$ analyses~\cite{Higgs:131007_coupling_comb,Higgs:131007_spin_comb}.
Events are selected using a loose diphoton trigger that is nearly 100\% efficient for events passing the offline photon selection.
Photons are reconstructed starting from clusters of energy deposited in the electromagnetic calorimeter. 
Events are required to contain two photon candidates whose calorimeter energy clusters match the expectations for photon-induced electromagnetic showers~\cite{ATL-PHYS-PUB-2011-007,Performance:120817_atlas_photon_id}.
The pseudorapidity \footnote{ATLAS uses a right-handed coordinate system with its origin at the nominal interaction point (IP) in the center of the detector and the $z$-axis along the beam pipe. The $x$-axis points from the IP to the center of the LHC ring, and the $y$-axis points upward. Cylindrical coordinates $(r,\phi)$ are used in the transverse plane, $\phi$ being the azimuthal angle around the beam pipe. The pseudorapidity is defined in terms of the polar angle $\theta$ as $\eta=-\ln\tan(\theta/2)$.},
$\eta$, of the two photons must fall within the geometric acceptance of the detector for photons, $|\eta| < 2.37$, 
  excluding the region between the barrel and endcap calorimeters ($1.37 < |\eta| < 1.56$).
The ratio of the transverse momentum of the leading (subleading) photon to the invariant mass of the pair, $\pT/\myy$, must exceed $0.35\,(0.25)$. The invariant mass of the pair is calculated 
as in Ref.~\cite{Higgs:131007_coupling_comb}.
Photons are required to be isolated:
  the energy in the calorimeter~\cite{Higgs:131007_coupling_comb,Aad:2010sp} within a cone of size
  $\Delta R \equiv \sqrt{\Delta \eta^2 + \Delta \phi^2} = 0.4$ 
  around the photon direction must be less than 6\,GeV, and the scalar sum of the $\pt$ of the tracks 
  in a cone of $\Delta R = 0.2$ must be less than 2.6\,GeV. In addition, 
  the photon pair must satisfy a broad requirement $105 < \myy < 160\gev$ for an event to be considered~\cite{Higgs:131007_coupling_comb,Higgs:131007_spin_comb}.

Jets are reconstructed from clusters of energy in the electromagnetic and hadronic calorimeters using the anti-$k_t$ algorithm~\cite{b-akt} 
  with a radius parameter of 0.4, starting from energy deposits grouped into 
  topological clusters~\cite{Barillari:1112035}. 
Simulation is used to correct jets for instrumental effects~\cite{jetcorr}
  and to account for the average energy in the detector in the event due to additional $pp$ interactions and the underlying event~\cite{Cacciari2008119}. 
The calibration is refined using \emph{in situ} measurements.
Jets are required to fall within the tracker acceptance of $|\eta| < 2.5$
  and satisfy $\pt > 35\gev$,
  with the leading jet in the event required to have $\pt > 55$\,GeV. 
Events with jets arising from noisy regions in the calorimeters, beam 
  backgrounds, or cosmic rays are rejected~\cite{jetcorr}.
Low-$\pt$ jets from additional proton--proton interactions in the same bunch crossing are rejected with a requirement on the scalar sum of the $\pt$ of tracks associated with the jet: 
  for jets with $|\eta| < 2.4$ and $\pt < 50\gev$, 
  tracks associated with the hard scatter vertex must contribute over 50\% of the sum.

Jets from the decay of long-lived heavy-flavor hadrons are selected 
using a multivariate tagging algorithm ($b$-tagging)~\cite{btagging2012} 
  with an efficiency of 70\% for jets from $b$-quark fragmentation in $\ttbar$ simulation.
The four-momenta of muons~\cite{ATLAS-CONF-2013-088} 
  closer than $\Delta R = 0.4$ to a $b$-tagged jet and with $\pt>4\gev$ are included in the jet four-momentum.

Events with at least two photons and two or more jets are selected for further analysis 
  if the invariant mass of the two leading jets is consistent with the decay of a Higgs boson. 
While the invariant mass resolution for the pairs of $b$-tagged jets is approximately $13\gev$,
  the mass window is chosen as $95 < m_{jj} <135\gev$ to account for the downward shift 
  of the mean from the true value due to effects such as unmeasured neutrinos from semi-leptonic $b$-decays.

In the non-resonance search, the background and potential signal are fit to the unbinned $\myy$ distribution
  of all events passing the dijet and diphoton selections described above.
This fit has three components: the signal with a pair of Higgs bosons,
  the background processes with a single Higgs boson resonant at $\myy=\mh$, 
  and the continuum background.
The single Higgs boson backgrounds are dominated by the processes with pairs of $b$-quarks, namely $\tth$ and $(Z\to b\bar b)h$, with smaller contributions from ggF, VBF, and $Wh$.
The combined acceptance and selection efficiency for the SM Higgs boson pair production signal is 7.4\%.
Simulation studies show that the continuum contribution in the signal region is split between events with two photons and events with a single photon in association with a jet faking the second photon.
The $b$-tagged jets include real heavy-flavor jets and mis-tagged light-flavor jets.
The contribution from dileptonic decays of $\ttbar$ events where two electrons fake the two photons 
  is roughly 10\% of the total background. 
The contribution from other processes is negligible.

The fit is performed simultaneously in two categories.
The first category is the signal region, in which at least two jets are $b$-tagged.
The second is a control region, containing events with fewer than two $b$-tags.
The two classes of events are kinematically identical:
  in the signal region, the mass and $\pt$ requirements defined above must be satisfied by the two leading tagged jets,
  whereas in the control region, they are met by the two leading jets.

Following earlier ATLAS analyses, the shape of the \myy{} resonance is described by the sum of a Crystal Ball function and a wide Gaussian component that models the tails of the distribution~\cite{Higgs:131007_coupling_comb}. 
An exponential function describes the continuum backgrounds that fall with \myy{}. 
The slope of the exponential is shared in the fit between the two categories so that the control region constrains the background shape in the signal region. Figure~\ref{fig:nonres} shows the separate diphoton mass distributions for events with $\geq 2\ b$-tags and events with $\leq 1\ b$-tag. 

\begin{figure}
\centering
\includegraphics[width=0.46\textwidth]{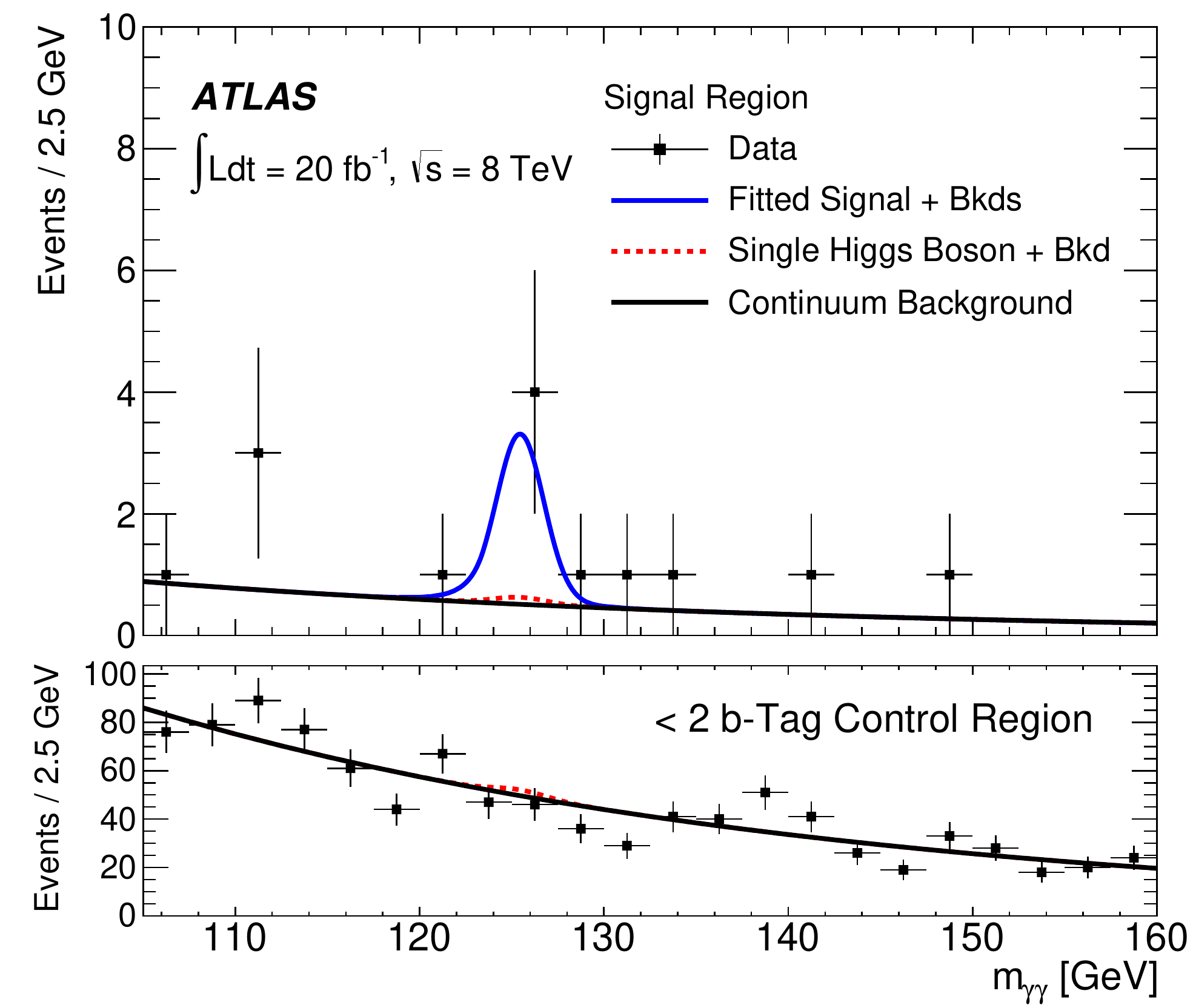}
\caption{
  (Upper plot) Diphoton invariant mass spectrum for data and the corresponding fitted signal and background in the signal region for the non-resonance search.
  (Lower plot) The diphoton invariant mass spectrum in the continuum background from events with fewer than two 
      $b$-tags and the corresponding fitted curve, the shape of which is also used in the upper plot.
}
\label{fig:nonres}
\end{figure}

The search for resonant production of pairs of Higgs bosons starts with the 
  same signal selection as above but imposes an additional requirement on \myybb. 
Due to the small number of expected events after this additional requirement, 
  the resonance analysis uses a counting experiment with cuts on \myy and \myybb, in place of the unbinned fit in \myy.
The cut on the diphoton mass is set as a window of twice the mass resolution, $\pm2\sigma_{\myy}$, around the Higgs boson mass, \mh = 125.5~\gev~\cite{Higgs:131007_coupling_comb}.
For this cut, the $\myy$ resolution is set to the expected value of $1.6\gev$.
The acceptance of this requirement on background events without Higgs bosons, $\epsilon_{\myy}$,
  is measured by fitting an exponential function to the \myy{} sidebands 
  for events with fewer than two $b$-tagged jets. 
For this fit, the $\myy$ region of $\mh\pm5\gev$ is excluded to eliminate any potential contamination from resonant Higgs boson production.
For $N$ observed events with two $b$-tags in the sideband ($|\myy -\mh| > 2\sigma_{\myy}$), 
  the number of expected non-Higgs boson background events ($N_{\myy})$ 
  within $2 \sigma_{\myy}$ around $m_h$ is given by:
\begin{align}
 N_{\myy} = N \frac{\epsilon_{\myy}}{1-\epsilon_{\myy}},
\end{align}
\noindent where the denominator compensates for the fact that $\epsilon_{\myy}=0.13$
  is derived relative to the full \myy{} spectrum while $N$ contains only those events in the sidebands.

Before reconstructing the four-object mass, $\myybb$, 
  a scaling factor of $\mh/\mbb$ is applied to the four-momentum of the $b\bar{b}$ system, where $\mh$ is set to the value of $125\gev$ used in simulation.
This improves the $\myybb$ resolution by 30--60\% depending on the mass hypothesis, without biasing or significantly altering the shape of the background.
Requirements are then made on $\myybb$ to select the smallest window containing 95\% of the previously-selected events,
  simulated for the narrow resonant signal hypotheses.
These requirements vary linearly with the mass, $m_X$, of the resonance considered. 
The width of the signal window varies from $17\gev$ at $m_X=260\gev$ to $60\gev$ at $m_X=500\gev$.
The acceptance for the continuum background to pass this requirement, $\epsilon_{\myybb}$, 
    also varies with $m_X$.
It is measured using events in data with $|\myy - \mh| < 2\sigma_{\myy}$
  and fewer than two $b$-tags. 
Studies in both data sidebands and simulation show that the shapes of $\myybb$ and $\myyjj$ agree within statistical uncertainties.
The distribution of $\myyjj$ in data is fitted with a Landau function, which is integrated in the signal window to obtain $\epsilon_{\myybb}$ for each mass hypothesis. 
The bottom panel of Fig.~\ref{fig:resonant} shows this fit.
The value of $\epsilon_{\myybb}$ is small $(<8\%)$ at low and high $m_X$, and peaks at 18\% for $m_X=300\gev$.
The combined acceptance and selection efficiency for a resonance signal to pass all requirements varies from 3.8\% at $m_X=260\gev$ to 8.2\% at $m_X=500\gev$.

\begin{figure}
\centering
\includegraphics[width=0.46\textwidth]{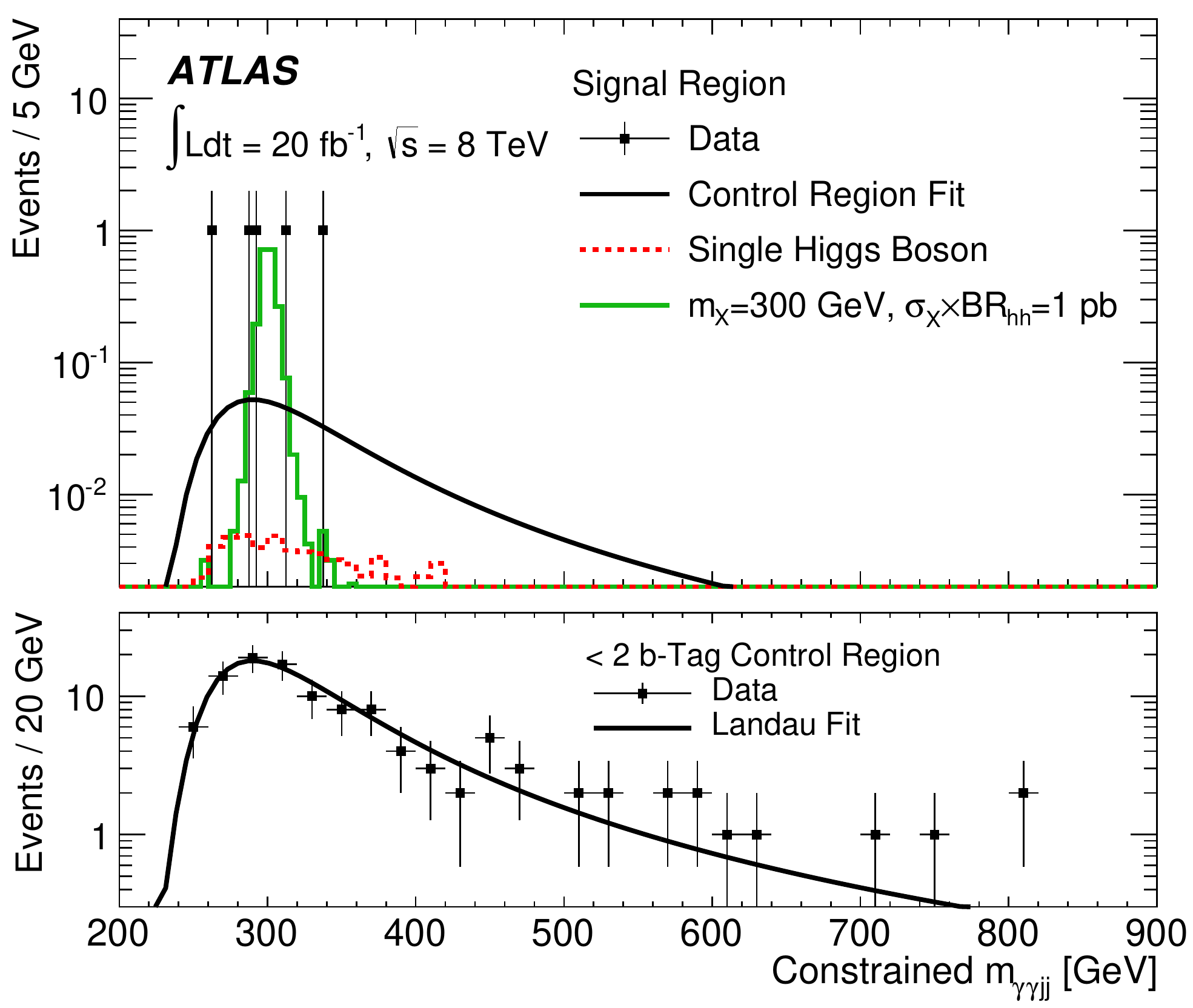}
\caption{
  (Upper plot) The constrained four-object invariant mass, $\myyjj$,
    for data events in the resonance signal region.
  The expected backgrounds are also shown.
  A narrow width resonance at 300 GeV is displayed for comparison only.
  (Lower plot) The diphoton invariant mass spectrum in the continuum background from events with fewer than two 
      $b$-tags and the corresponding fitted curve, the shape of which is also used in the upper plot.
}
\label{fig:resonant}
\end{figure}

The total background from sources without Higgs boson decays in the resonance analysis $N_{\rm B}$ is given by:
\begin{align}
N_{\rm B} = N\frac{\epsilon_{\myy}}{1-\epsilon_{\myy}}\epsilon_{\myybb},
\end{align}
\noindent
where $N$ is the number of events in the $\myy$ sidebands, and $N_{\rm B}$ and $\epsilon_{\myybb}$ are functions of $m_X$. 
Uncertainties on this extrapolation are described below.

Because they are not accounted for by the above $\myy$ sideband techniques, 
  contributions from single Higgs bosons produced in association with jets (particularly with $c\bar c$ or $\bbbar$ pairs) are estimated using simulation.  
In the resonance analysis, the yield from the non-resonant SM $hh$ processes is similarly included.  
SM cross sections and branching fractions are assumed in all cases~\cite{Higgs:130704_YR_3_prop}.

Most systematic uncertainties are small when compared to statistical uncertainties, in particular for the resonance search. 
  
The evaluation of experimental uncertainties on photon identification (2.4\%) and isolation efficiencies (2\%) 
  follow the methods used in the inclusive ATLAS $\hyy$ analyses~\cite{Higgs:131007_coupling_comb,Higgs:131007_spin_comb,Higgs:130704_YR_3_prop}. The theoretical uncertainties on the single Higgs boson backgrounds are similarly adopted.
Because there are no heavy flavor quarks at lowest order associated with ggF or VBF production, additional uncertainties are evaluated for these higher-order processes.  These uncertainties are derived from a comparison of simulated predictions to data for similar initial states: gluon-initiated production of $\ttbar$ with heavy flavor~\cite{Aad:2013tua} for ggF, and quark-initiated $W$ boson production with heavy flavor~\cite{Aad:2013vka} for VBF.  Since the ggF and VBF contributions are less than 15\% of the expected single Higgs boson yield in the signal region, the net impact of these uncertainties remains small. PDF and scale uncertainties on SM $hh$ production are taken from Ref.~\cite{Dihiggs:121221_djouadi_self-coupling}.

Because of the cuts on the ratio $\pT/\myy$,
  photon energy scale uncertainties are negligible.
The uncertainty of 13\% on the diphoton mass resolution $\sigma_{\myy}$ is propagated into the resonance search as a 1.6\% uncertainty on the number of events migrating into and out of the signal region.
This represents the fraction of events where an upward variation of the photon resolution
  causes the diphoton mass to leave the $\mh\pm2\sigma_{\myy}$ window required for the signal region.
The uncertainty on $m_h$
  impacts the peak position in \myy{} in the signal plus background fit of the non-resonance analysis, 
  and in the resonance search it is transformed into a 1.7\% uncertainty on the number of signal events in the mass window.
The uncertainty for the acceptance of the $\myy$ cuts on
  non-Higgs boson backgrounds is estimated by comparing fits of $\myy$
  to data in control regions with reversed photon identification or $b$-tagging requirements,
  and using different functional forms.
The largest deviation observed from these fits (11\%) is used for all searches. 

Three components contribute to the uncertainty on $\epsilon_{\myybb}$, 
  and are combined in quadrature.
(1) The limited number of events in the control region with fewer than two $b$-tags used for the Landau fit lead to a
      relative statistical uncertainty of 3--18\% that varies as a function of $m_X$.
(2) The $\myyjj$ shape for untagged jets might not exactly mirror the one for tagged jets.
    The tagged and untagged samples are compared in simulation and 
      the relative difference in $\epsilon_{\myybb}$ is taken as the uncertainty.
    This value varies with $m_X$ and is always less than 30\%.
(3) Finally, an uncertainty of 16--30\%, depending on $m_X$,
    is included to cover the choice of the analytic function.  This was evaluated 
    via comparisons of Landau shapes to alternate functions in simulation, including Landau shapes where the width varies with $\myybb$, as well as Crystal Ball functions.
Potential contamination from SM single Higgs boson processes in the control region
  is estimated to be less than 4\% and is subtracted with negligible impact on the shape.

Uncertainties due to the $b$-tagging calibration are typically 
  2--4\% for both the single Higgs boson and signal processes. 
Uncertainties due to the jet energy scale are 7\% (22\%) 
  for single Higgs boson backgrounds in the non-resonance (resonant) analysis, 
  and 1.4\% (4.4\%) for signal processes. 
Uncertainties due to the jet energy resolution are 4.8\% (21\%) 
  for single Higgs boson backgrounds, and 6.3\% (9.3\%) for signal processes. The uncertainty on the integrated luminosity is 2.8\%. It is derived, following the same methodology as that detailed in Ref.~\cite{Detectors:130218_atlas_lumi}, from a preliminary calibration of the luminosity scale derived from beam-separation scans performed in November 2012.

The combined, unbinned signal plus background fit for the non-resonance analysis is shown in Fig.~\ref{fig:nonres}.
Within a $\pm2\sigma_{\myy}$ window around the Higgs boson mass, 
  1.5 events are expected, with $1.3\pm0.5$ from the continuum background 
  and $0.17\pm0.04$ from single Higgs boson production, which is dominated by $\tth$ events. About 0.04 events are expected from SM Higgs boson pair production.
Five events are observed, corresponding to $2.4\sigma$ from the background-only hypothesis,
  using the test statistic based on the profile likelihood ratio \cite{asym} with the hypothesized signal rate set to zero.
The 95\% confidence level ($CL$) upper limit on the Higgs boson pair production cross section
  is calculated using the frequentist $CL_S$ method~\cite{Statistics:020716_read_cls}. 
Exclusions and significances are evaluated using pseudo-experiments.
Assuming SM branching ratios for the light Higgs boson decays, 
  the expected upper limit is $1.0_{-0.2}^{+0.5}\pb$; 
  the observed limit is $2.2\pb$.

For the resonance analysis, as before, SM branching fractions
for the light Higgs boson are assumed.
The expected exclusion improves from 1.7 to $0.7 \pb$ 
  as a function of $m_X$ from 260 to $500\gev$, as shown in Fig.~\ref{fig:exclusion}. 
This behavior derives from increased event-level acceptance at larger masses. The observed exclusion ranges from 3.5 to $0.7 \pb$.
The five events selected in the $\myy$ signal region are shown in $\myybb$, in Fig.~\ref{fig:resonant}.
The local probability of compatibility to the background-only hypothesis, $p_0$, 
  reaches a minimum of $0.002$ at $m_X = 300\gev$, 
  corresponding to $3.0\sigma$.
The number of events lying within the $\myybb$ window of each mass hypothesis is readily apparent in `steps' in the exclusion plot.
The step size used for the limit is reduced in the range near the observed events, to show this structure.
A look-elsewhere effect~\cite{lee,Statistics:100601_gross_LEE} is evaluated by generating pseudo-datasets of the background-only hypothesis, and identifying the mass hypothesis with the lowest $p$-value in each.
The global probability of an exccess as significant as the observation to occur at any mass in the range studied is found to be 0.019, corresponding to $2.1\sigma$.

The limits derived are juxtaposed in Fig.~\ref{fig:exclusion}
with the prediction for an illustrative type I 2HDM~\cite{Branco:2011iw, Eriksson:2009ws, Harlander:2012pb} not excluded by current data with
$\cos(\beta-\alpha)=-0.05$ and $\tan(\beta)=1$. 
The heavy Higgs bosons are taken to be degenerate in mass,
    and the mass of the lightest CP-even Higgs boson is set to $125\gev$. 
All major production mechanisms of $H\rightarrow hh$ are considered.
Cross sections and branching ratios were calculated as
  described in Ref.~\cite{arxiv:1312.5571}.

\begin{figure}
\centering
\includegraphics[width=0.47\textwidth]{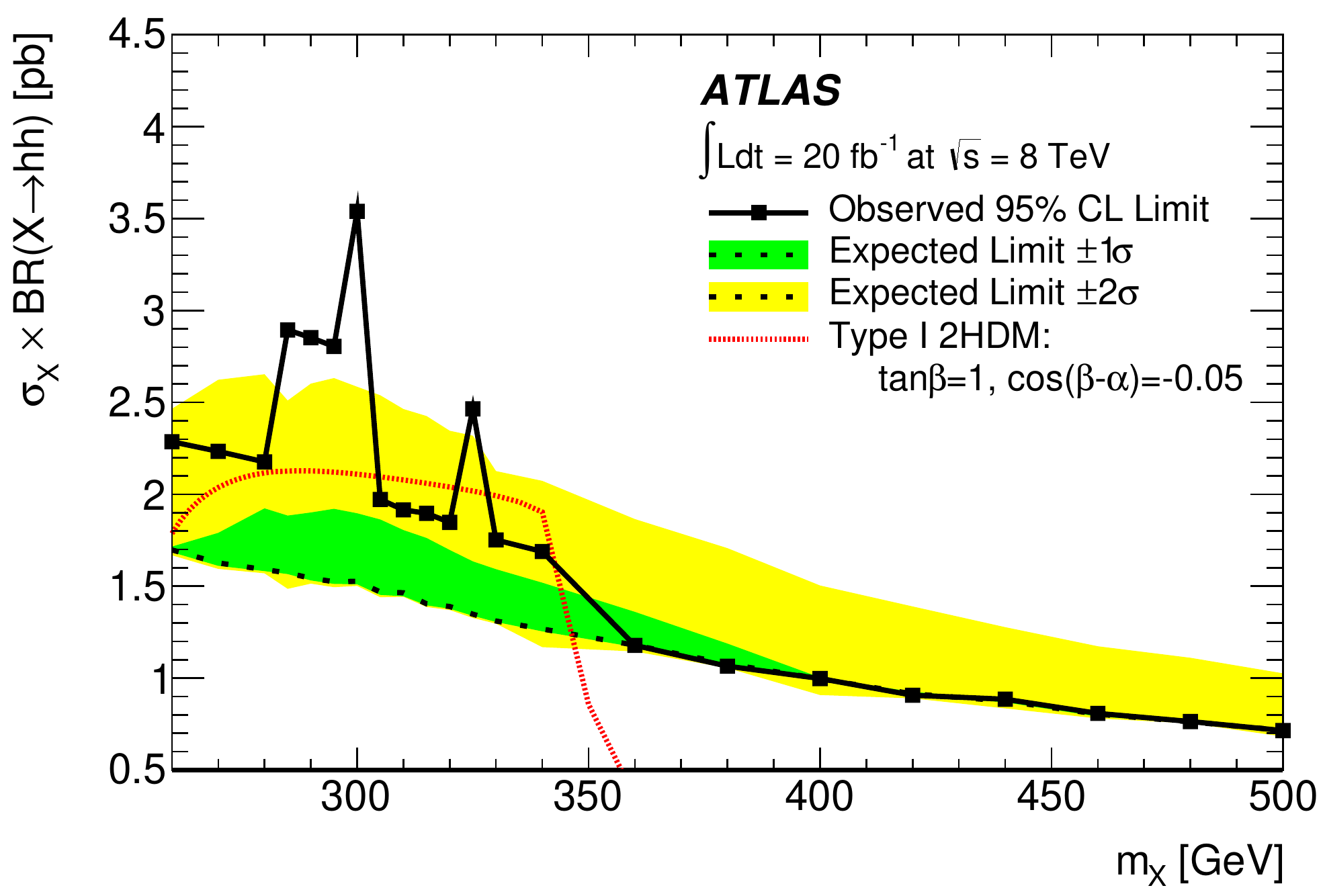}
\caption{
  A 95\% $CL$ upper limit on the cross section times branching ratio of 
    a narrow resonance decaying to pairs of Higgs bosons as a function of $m_X$
  (see text for more details).
}
\label{fig:exclusion}
\end{figure}

In conclusion, this Letter presents searches for resonant and non-resonant Higgs boson pair production
  using $20.3\ifb$ of proton--proton collision data at $\sqrt{s} = 8\tev$
  generated by the Large Hadron Collider and recorded by the ATLAS detector in 2012.
A 95\% confidence level upper limit is placed on the non-resonant production cross section at $2.2\pb$, 
  while the expected limit is $1.0_{-0.2}^{+0.5}\pb$. 
The difference derives from a small excess of events, corresponding to $2.4\sigma$.

In the search for a narrow resonance decaying to a pair of Higgs bosons, 
   the expected exclusion on the production cross section falls from 
  $1.7\pb$ for a resonance at $260\gev$ to $0.7\pb$ at $500\gev$.
The observed exclusion ranges from 0.7--3.5$\pb$. 
It is weaker than expected for resonances below 350~GeV.

We thank CERN for the very successful operation of the LHC, as well as the
support staff from our institutions without whom ATLAS could not be
operated efficiently.

We acknowledge the support of ANPCyT, Argentina; YerPhI, Armenia; ARC,
Australia; BMWFW and FWF, Austria; ANAS, Azerbaijan; SSTC, Belarus; CNPq and FAPESP,
Brazil; NSERC, NRC and CFI, Canada; CERN; CONICYT, Chile; CAS, MOST and NSFC,
China; COLCIENCIAS, Colombia; MSMT CR, MPO CR and VSC CR, Czech Republic;
DNRF, DNSRC and Lundbeck Foundation, Denmark; EPLANET, ERC and NSRF, European Union;
IN2P3-CNRS, CEA-DSM/IRFU, France; GNSF, Georgia; BMBF, DFG, HGF, MPG and AvH
Foundation, Germany; GSRT and NSRF, Greece; ISF, MINERVA, GIF, I-CORE and Benoziyo Center,
Israel; INFN, Italy; MEXT and JSPS, Japan; CNRST, Morocco; FOM and NWO,
Netherlands; BRF and RCN, Norway; MNiSW and NCN, Poland; GRICES and FCT, Portugal; MNE/IFA, Romania; MES of Russia and ROSATOM, Russian Federation; JINR; MSTD,
Serbia; MSSR, Slovakia; ARRS and MIZ\v{S}, Slovenia; DST/NRF, South Africa;
MINECO, Spain; SRC and Wallenberg Foundation, Sweden; SER, SNSF and Cantons of
Bern and Geneva, Switzerland; NSC, Taiwan; TAEK, Turkey; STFC, the Royal
Society and Leverhulme Trust, United Kingdom; DOE and NSF, United States of
America.

The crucial computing support from all WLCG partners is acknowledged
gratefully, in particular from CERN and the ATLAS Tier-1 facilities at
TRIUMF (Canada), NDGF (Denmark, Norway, Sweden), CC-IN2P3 (France),
KIT/GridKA (Germany), INFN-CNAF (Italy), NL-T1 (Netherlands), PIC (Spain),
ASGC (Taiwan), RAL (UK) and BNL (USA) and in the Tier-2 facilities
worldwide.

\bibliographystyle{apsrev4-1}
\bibliography{paper}
\onecolumngrid 
\clearpage
\begin{flushleft}
{\Large The ATLAS Collaboration}

\bigskip

G.~Aad$^{\rm 84}$,
B.~Abbott$^{\rm 112}$,
J.~Abdallah$^{\rm 152}$,
S.~Abdel~Khalek$^{\rm 116}$,
O.~Abdinov$^{\rm 11}$,
R.~Aben$^{\rm 106}$,
B.~Abi$^{\rm 113}$,
M.~Abolins$^{\rm 89}$,
O.S.~AbouZeid$^{\rm 159}$,
H.~Abramowicz$^{\rm 154}$
% ...
\bigskip
\\
% ...
$^{11}$ Institute of Physics, Azerbaijan Academy of Sciences, Baku, Azerbaijan\\
% ...
$^{84}$ CPPM, Aix-Marseille Universit{\'e} and CNRS/IN2P3, Marseille, France\\
% ...
$^{89}$ Department of Physics and Astronomy, Michigan State University, East Lansing MI, United States of America\\
% ...
$^{106}$ Nikhef National Institute for Subatomic Physics and University of Amsterdam, Amsterdam, Netherlands\\
% ...
$^{112}$ Homer L. Dodge Department of Physics and Astronomy, University of Oklahoma, Norman OK, United States of America\\
$^{113}$ Department of Physics, Oklahoma State University, Stillwater OK, United States of America\\
% ...
$^{116}$ LAL, Universit{\'e} Paris-Sud and CNRS/IN2P3, Orsay, France\\
% ...
$^{152}$ Institute of Physics, Academia Sinica, Taipei, Taiwan\\
% ...
$^{154}$ Raymond and Beverly Sackler School of Physics and Astronomy, Tel Aviv University, Tel Aviv, Israel\\
% ...
$^{159}$ Department of Physics, University of Toronto, Toronto ON, Canada\\
% ...
\end{flushleft}

\end{document}